\documentclass[final,3p,times,authoryear]{elsarticle}

\usepackage{amssymb}
\usepackage{lipsum}
\usepackage{url}
\usepackage{amsmath}
\usepackage[normalem]{ulem}

\usepackage{lineno}

\journal{Advances in Space Research}

\begin{document}

\begin{frontmatter}



\title{The Random Hivemind: An Ensemble Deep Learner Application to Solar Energetic Particle Prediction Problem}


\author[1]{Patrick M. O’Keefe}
\address[1]{{Computer Science Department, New Jersey Institute of Technology},
            323 Dr Martin Luther King Jr Blvd, 
            Newark
            07102, 
            NJ,
            USA}

\author[2]{Viacheslav Sadykov}
\address[2]{{Physics \& Astronomy Department, Georgia State University},
            25 Park Place NE, 
            Atlanta
            30303, 
            GA,
            USA}

\author[3,4]{Alexander Kosovichev}
\address[3]{{Physics Department, New Jersey Institute of Technology},
            323 Dr Martin Luther King Jr Blvd, 
            Newark
            07102, 
            NJ,
            USA}
\address[4]{{NASA Ames Research Center},
            NASA Research Park, 
            Moffett Field
            94035, 
            CA,
            USA}

\author[4]{Irina N. Kitiashvili}

\author[1]{Vincent Oria}

\author[3]{Gelu M. Nita}

\author[1]{Fraila Francis}

\author[1]{Chun-Jie Chong}

\author[3]{Paul Kosovich}

\author[2]{Aatiya Ali}

\author[5,2]{Russell D. Marroquin}
\address[5]{{Department of Physics, University of California San Diego},
            9500 Gilman Dr, 
            La Jolla
            92093, 
            CA,
            USA}

\begin{abstract}
The application of machine learning and deep learning techniques, including the wide use of non-ensemble, conventional neural networks (CoNN), for predicting various phenomena has become very popular in recent years thanks to the efficiencies and the abilities of these techniques to find relationships in data without human intervention. However, certain CoNN setups may not work on some datasets, especially if the parameters passed to it, including model parameters and hyperparameters, are arguably arbitrary in nature and need to continuously be updated with the need to retrain the model, especially if the additions of new features render old models obsolete. This concern can be partially alleviated by employing committees of neural networks that are identical in terms of input features and initialized randomly and ``vote'' on the decisions made by the committees as a whole. Yet, members of the committee have similar architectures and features passed to them, making it possible for the committee members to ``agree'' on identical sets of weights and biases for all nodes and edges. Members of these committees also cannot be expanded to accommodate new features and entire committees must therefore be retrained in order to do so. We propose the Random Hivemind (RH) approach, which helps to alleviate this concern by having multiple neural network estimators make decisions based on random permutations of features and prescribing a method to determine the weight of the decision of each individual estimator. The effectiveness of RH is demonstrated through experimentation in the predictions of hazardous Solar Energetic Particle (SEP) events by comparing it to that of using both CoNNs and the aforementioned setup of committees identical in input features in this application. Our results demonstrate that RH, while having a comparable or better performance than the CoNN and a Committee-based approach, demonstrates a lesser score spread for the individual experiments, and shows promising results with respect to capturing almost every single flare instance leading to SEPs.
\end{abstract}




\end{frontmatter}
\tableofcontents


\section{Introduction} \label{sec:intro}

    The prediction of Solar Energetic Particle (SEP) events and the understanding of their precursors represent major challenges in heliophysics and space weather from both the operational and the research perspectives. Increased fluxes of SEPs are of interest to various users, from governmental and private space weather agencies to airlines and power grid operators. Routine daily forecasting and short-term warning and alert systems for Solar Proton Events (SPEs) were implemented by the Space Weather Prediction Center (SWPC) at the National Oceanic and Atmospheric Administration \citep[NOAA,][]{balch1999,balch2008}. The performance of the operational forecasting systems is still far from predicting every single SEP event \citep{bain2021}.
    
    SEP events are initiated by solar flares and coronal mass ejections \citep[CMEs,][]{reames2021}. Statistical relations between the flare soft X-ray properties \citep[such as the peak ratios of the 1-8$\AA$ and 0.5-4$\AA$ fluxes, which is proportional to the flare temperature computed in a single-temperature approximation,][]{ryan2012,sadykov2019} and the consequent CMEs and SEPs have been known for a long time. Solar flares are classified in terms of the maximum soft X-ray (SXR) flux observed by the NOAA Geostationary Operational Environmental Satellite (GOES) Network in the 1-8$\AA$ wavelength range. In this classification, the A-class flares have the maximum soft X-ray flux greater than $10^{-8}$ W/m$^2$; for the B, C, M, and X-class flares, the SXR is greater than $10^{-7}$, $10^{-6}$, $10^{-5}$, and $10^{-4}$ W/m$^2$ correspondingly. In particular, it was found that the lower the soft X-ray class, the greater the difference in the peak temperatures between the SEP-associated and non-SEP flares, with lower temperatures corresponding to the SEP-associated flares \citep{garcia1994}. These relations were quantified and utilized for forecasting SEPs using a larger number of flare events \citep{garcia2004}. The results were also reproduced later \citep{kahler2018}, where the authors attempted to predict the SEP-associated flares using the k-nearest neighbors machine learning algorithm and neural networks separately for the Western and Eastern hemispheres of the Sun. Although the observed relationships are clear, the exact reason why the lower-temperature flares are more associated with SEPs remains largely unknown \citep{kahler2018}. In addition, the durations and temperatures of the flares were found to be related statistically to the properties of CMEs \citep{ling2020,kahler2022}. Thus, the flare duration and temperature can be used to constrain SEP parameters \citep{kahler2013} or serve as a basis for establishing empirical models for SEP forecasting \citep{richardson2018}. The Empirical model for Solar Proton Event Real-Time Alert (ESPERTA) forecasting tool \citep{Laurenza2009,Laurenza2018} also utilizes the integrated SXR intensities and integrated radio intensities at 1\,MHz to provide short-term predictions of SPEs with proton energy higher than 10\,MeV and 100\,MeV.
    
    The extension of these works is the employment of machine learning (ML) and deep learning techniques, in particular, for forecasting SEPs based on the properties of the preceding (parental) solar flares. For example, \citet{Aminalragia2021} employed neural networks trained on the time series of the SXR fluxes observed during the solar flares directly. The authors found that the model can predict a large majority of SEP-associated flares (higher than 85\%) during the period of 1988-2013 while maintaining a low false-positive rate. \citet{Boubrahimi2017} analyzed the correlations among the GOES soft X-ray and proton flux time series and employed the classification decision tree model for predicting the 100\,MeV SPEs. \citet{Lavasa2021} analyzed a variety of ML algorithms such as random forest, neural networks, extremely randomized trees, and extreme gradient boosting. They concluded that, among the soft X-ray parameters, fluence is the most important for predicting SEPs.

    The utilization of Conventional Neural Networks (hereafter CoNNs, defined in this work as a single, i.e., non-ensemble, neural network of any architecture) for space weather prediction purposes is very common in literature \citep[e.g.,][etc.]{Torres2022,Nishizuka2020}. While CoNNs are very flexible and malleable in how they train on new data, the parameters they provide for the aspects, such as the size and shape of a given model, hyperparameters, and model selection, may need to be adjusted to be used on other data sets. This comes with the consequence of a requirement to continuously retrain models as data becomes increasingly outdated or if new features are to be added. A model's number of epochs and learning rate ideally need to be also adjusted based on the features fed into the model. In particular, the number of epochs low enough and the learning rate high enough can help to avoid overfitting the model to noise data \citep{afaq2020,you2019}, yet one needs to make sure that the model is trained to a sufficient extent in general to learn the patterns in the data. 
    This conflict may lead to a particular combination of an epoch count and learning rate for an entire model being sub-optimal, as a single combination may not work for all features, especially if some are more deterministic of the true labels of a given data set than others. Compromises could be made to adjust these parameters, but said compromises may, again, cause a given model to underfit based on some parameters and overfit based on others. One of the strategies to counter overfitting, in general, is to employ ensemble learning \citep{cunningham2000}.

    Another challenge related to the SEP prediction is that these events are rare and will represent minority-class events for the classification problem. For example, the ratio of the number of days with the enhanced proton flux with the energy greater than 10\,MeV and the particle flux greater than 10\,pfu (one particle flux unit, pfu, is equal to one particle per cm$^2$ per second per steradian) to the number of days with no enhanced flux is $\sim{}1/23$ for Solar Cycles 22-24 \citep{ali2023}. This ratio is even smaller ($\sim{}1/34$) for Solar Cycle 24 alone \citep{sadykov2021}. It was concluded for the ESPERTA model~\citep{Stumpo2021} that the performance of the algorithm (specifically, the False Alarm Rate, FAR) depends on the class-imbalance ratio in the train data set. Various techniques can be implemented to deal with class-imbalanced data, such as oversampling, undersampling, and misclassification weights \citep{ahmadzadeh2021}, and using synthetic data \citep{chen2021}. In addition to the traditional data-centric approaches to dealing with class imbalance, ensemble classifiers can be employed in such problems \citep{galar2012}. With respect to the problem of the prediction of SEPs, promising results were previously obtained employing neural network-based Committee ensembles \citep{Aminalragia2021} and random forest ensemble algorithms \citep{Lavasa2021}.
    
    In our previous work, we presented the application of the random forest ML algorithm for predicting SEPs and tested various class-imbalance treatment techniques \citep{okeefe2022}. In this work, we expand our investigation to new types of ML algorithms, including Conventional Neural Networks (CoNN), an ensemble of CoNNs following a voting approach \citep[Committee,][]{Aminalragia2021}, and introduce a weighted consensus that we call a Random Hivemind (RH). Both considered ensemble approaches are so-called ``bagging'' ensemble classifiers when individual ensemble members do not depend on each other and deterministically contribute to the classification decision. \textit{Investigation of the relative performance of the algorithms on the given data set of flares associated with SEPs is the primary focus of this paper.} The paper is structured as follows. Section~\ref{sec:datapreparation} describes the data preparation employed in this paper, namely the processing of the soft X-ray data, the association of flares and SEPs, and the preparation of data sets ready for machine learning (ML) analyses. Section~\ref{sec:mlmethodology} describes the ML algorithms tested in this work. The results and discussion are presented in Section~\ref{sec:results} and followed by the conclusion in Section~\ref{sec:discussion}.

\section{Data Preparation}
    \label{sec:datapreparation}

    The solar soft X-ray (SXR) emission observed by the GOES satellites in 0.5-4\,$\AA$ and 1-8\,$\AA$ wavelength channels can be represented under a single-temperature plasma approximation by two parameters, namely the plasma temperature ($T$) and its emission measure ($EM$). We utilize the legacy data set of the $T$ and $EM$ values estimated using the Temperature and Emission Measure Based Background Subtraction algorithm \citep[TEBBS,][]{ryan2012,sadykov2017} and collected in the Interactive Multi-Instrument Database of Solar Flares \citep[IMIDSF\footnote{\url{https://data.nas.nasa.gov/helio/portals/solarflares/}},][]{sadykov2017} for the 2002-2017 time period. In addition to peak values of the temperature and emission measure, $T_{max}$ and $EM_{max}$, we utilize the background-subtracted flare classes ($SXR_{max}$), flare durations, the times of the peaks of $T_{max}$, $EM_{max}$, and $SXR_{max}$ relative to the flare start and end times, and the observed disk X- and Y-coordinates of the host flare. In total, this data processing gives 12 parameters for every solar flare. For some flares, due to the complexity of the X-ray emission variations and high background noise, this algorithm failed (producing unrealistic $T_{max} \geq 100$\,MK, and negative time differences), and such flares were excluded from our data set. The total number of flares included in our analysis is 18311. Among these are: 5 A-class, 6919 B-class, 10074 C-class, 1207 M-class, and 106 X-class flares according to the SXR classification. These flare classes were calculated after subtracting the SXR background. As illustrated in Figure~4c of \citet{sadykov2019}, the background subtraction mostly affects the flare class of the weak BC-class flares, yet leading to more reliable behaviors of T and EM curves.

    \begin{figure}[ht]
        \centering
        \includegraphics[width=0.75\linewidth]{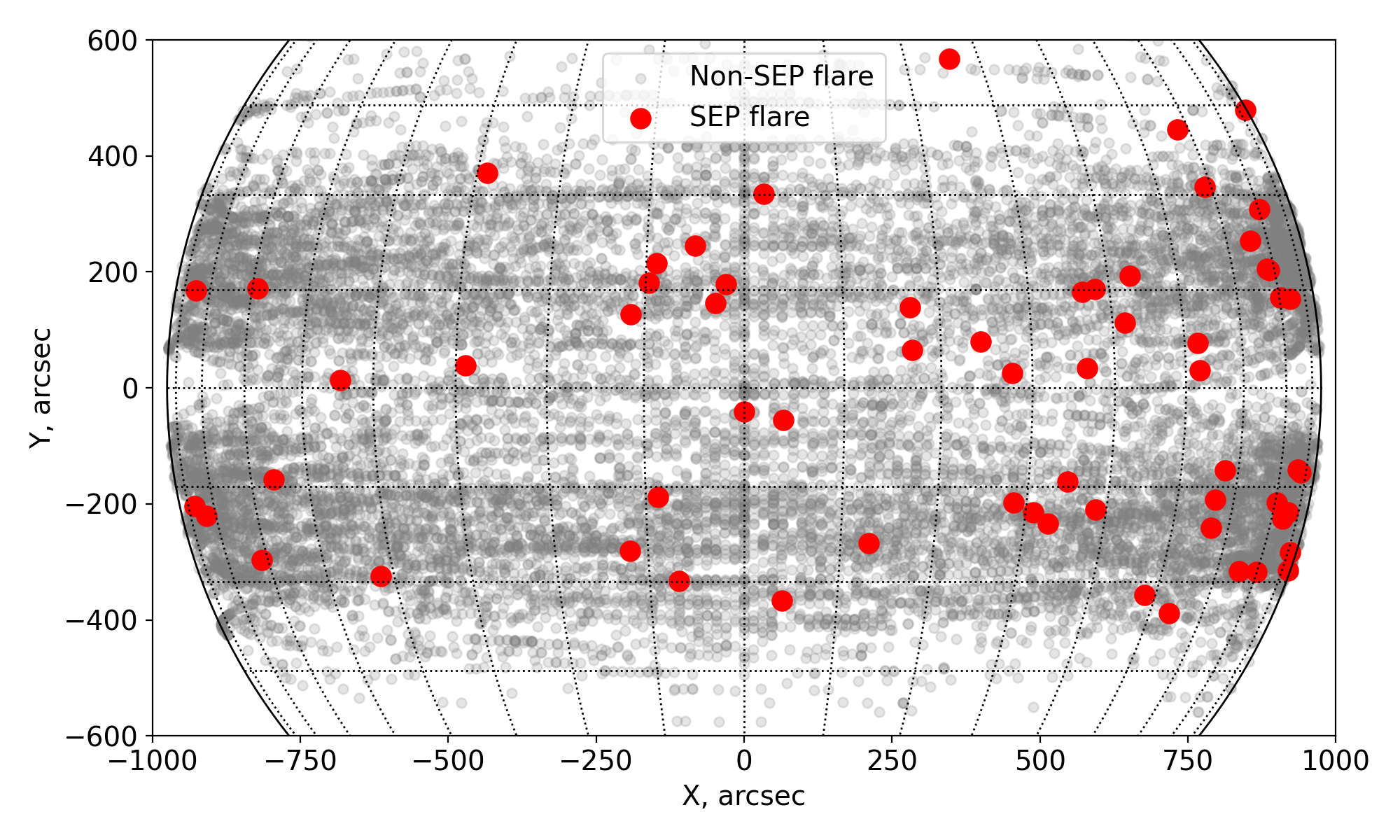}
        \caption{The locations of the SEP and non-SEP flares in the data set on the solar disk. Gray dots mark the solar flares that did not produce SEP events, and red dots mark the flares that resulted in the SEP events.}
        \label{fig:locations}
    \end{figure}
    
    To associate the flares with the SEP records, we utilize the list of the Solar Proton Events Affecting the Earth Environment\footnote{\url{https://umbra.nascom.nasa.gov/SEP/}} provided by the NOAA Space Environment Services Center. This data set represents the SEP events with the peak flux of $>$10\,MeV protons higher than 10\,pfu. A total of 64 flares from our list were associated with the SEP events and 18247 flares were without SEPs (non-SEP flares), providing an extreme class-imbalance ratio of $~$1/285. In terms of the $SXR_{max}$ parameter, 8 of the SEPs correspond to the C-class flares, 36~--- to M-class, and 20~--- to X-class flares. The list of the studied flares is publicly available at the Solar Energetic Particle Prediction Portal (SEP$^3$) website\footnote{\url{https://sun.njit.edu/SEP3/datasets.html}}. We note here, that the number of SEP events considered in some other works on the SEP prediction problem \citep[e.g.,][]{Papaioannou2016,Lavasa2021} was significantly larger (314 and 257, correspondingly) than in the current work. Both of these studies consider a longer time span (events detected during 1984-2013 and 1988-2013) which led to larger statistics of SEP events. Also, in the current work, we omit from consideration the SEP events which (1) do not have an association with the flare event, and (2) do not have the coordinates of the host flare event identified according to the GOES flare catalog. Both restrictions are leading to the loss of 15 SEP event records, leaving us with 64 events in total. Figure~\ref{fig:locations} illustrates the locations of the SEP and non-SEP flares on the solar disk. It is evident that the SEP flares have a preference to originate in the Western hemisphere, which has a more direct magnetic connectivity to Earth, although some SEP flares have originated close to the Eastern limb as well.
        
    \begin{figure}[ht]
        \centering
        \includegraphics[width=1.0\linewidth]{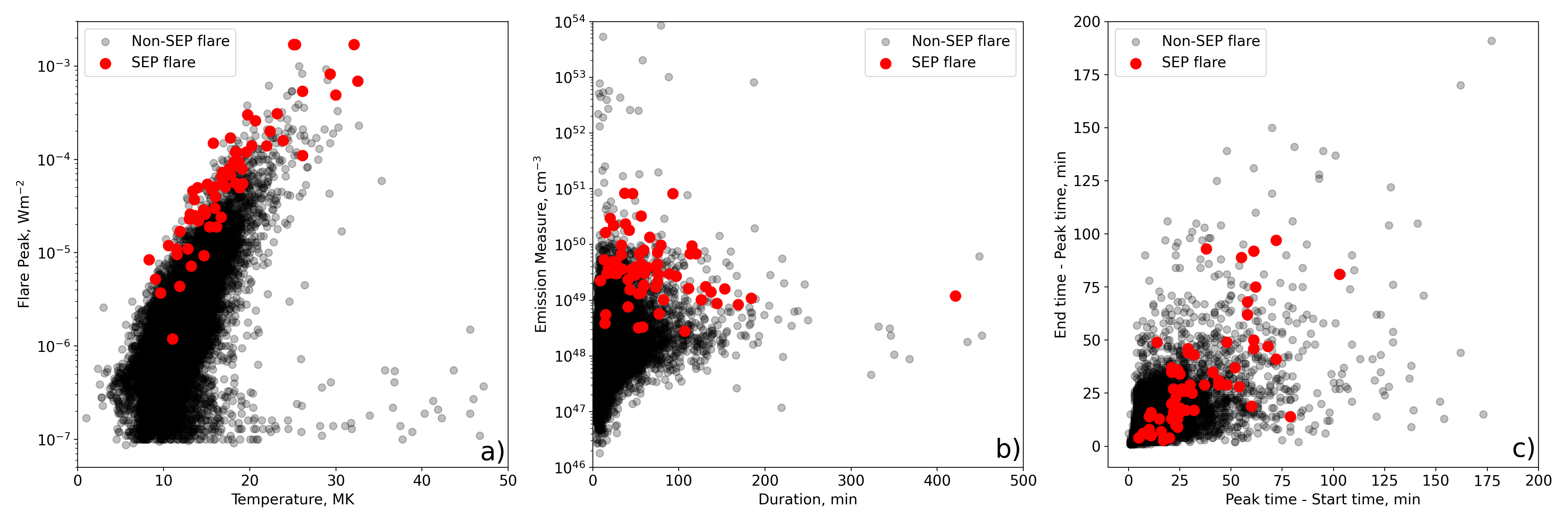}
        \caption{(a) Distribution of the SEP and non-SEP flares in the soft X-ray flux at 1-8\,$\AA$ and flare peak temperatures diagram, (b) the flare duration and the peak emission measure diagram, and (c) the flare rise time and flare decay time diagram. Black dots mark the solar flares that did not produce SEP events, and red dots mark the flares that resulted in the SEP events.}
        \label{fig:dataset}
    \end{figure}

    The distribution of the SEP-associated and non-SEP solar flares in the diagrams of the flare temperature vs. the soft X-ray peak flux, the flare duration vs. the emission measure, and the flare rise time vs. the flare decay time are presented in Figure~\ref{fig:dataset}. Here, the flare rise time is determined as the time of $SXR_{max}$ minus the flare start time, and the flare decay time is determined as the flare end time minus the time of $SXR_{max}$. One can see that the flares that resulted in SEPs are not distributed randomly, even among the flares of the same SXR peak fluxes. Specifically, Figure~\ref{fig:dataset}a indicates that SEP-associated flares are colder on average among the flares with the same SXR peak flux (or the flare class). A similar dependence was found by \citet{garcia1994,garcia2004}. While no obvious pattern is observed for the parameters presented in Figures~\ref{fig:dataset}b~and~\ref{fig:dataset}c, a more detailed investigation is required to understand their relations to SEP-associated flares. For the ML analysis, we subdivide the data set into training and testing subsets; the latter contains 30\% of the full dataset. The train-test separation is randomly repeated ten times for every machine-learning experiment presented in this paper.

\section{Machine Learning Methodology}
    \label{sec:mlmethodology}
    
    Three neural network-based approaches are considered in this paper for the problem of prediction of SEP events. The first is the conventional neural network (hereafter CoNN), which represents a single fully connected neural network architecture. For ensemble deep learners, two more neural network-based approaches are constructed. The first ensemble approach is the Committee scheme employed by \citet{Aminalragia2021}, which involves several neural network estimators with the same input features and input layer shapes. The second ensemble approach is a Random Hivemind (RH) built using random down-selection of the flare characteristics from the training data as input features.

    There are several motivations to consider the development of the RH approach. First, the RH allows for individual estimators to be grown individually to accommodate additional features without the need to retrain the entire ensemble. By requiring some, but not all, of the estimators to be retrained, this may reduce the amount of time it takes to retrain deep learning models. Second, the RH not only takes advantage of the ability of ensemble learning to counter overfitting in general \citep{cunningham2000}, but also adjusts the training epoch counts, learning rates, and voting weights of individual estimators depending on the importance of the features used in these estimators. This allows RH to accommodate the risk of overfitting by reducing the chances of less important features being included or influencing the prediction.
    
\begin{figure}[ht]
        \centering
        \includegraphics[width=1.0\linewidth]{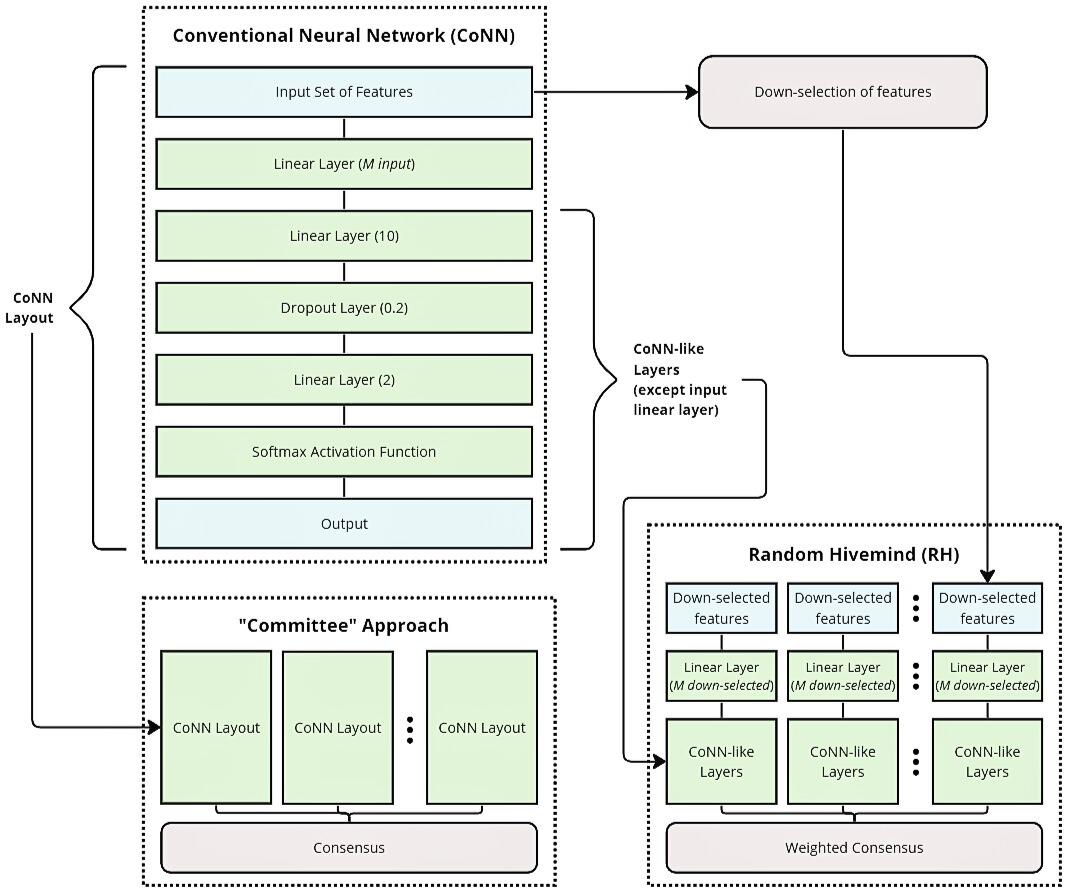}
        \caption{The schematic representation of the layouts of the Conventional Neural Network (CoNN), Committee Network, and Random Hivemind Network (RH). The numbers in parentheses for the linear layers indicate the number of neurons in the layer, and the number in parentheses for the dropout layer indicates the probability of each connection/weight being dropped from the training procedure.}
        \label{fig:network}
    \end{figure}

    For the series of tests presented in this paper, we consider two RH realizations that have several differences. First, the realizations are using a different number of input features. The first RH implementation (hereafter RH.v1) uses the square root of the total number of features, rounded up, as the number of input features for each neural network within an RH. The second implementation (hereafter RH~v2) uses half of the features as input. The layout between estimators remains unchanged within each Committee. In some sense, the semi-random selection of features (the probability of selection is yet proportional to their weight; see below) is inspired by the Random Forest ensemble learning algorithm \citep{Brieman2001RF}. Correspondingly, one has 12 features entering the CoNN or each committee member, and 4 or 6 features selected using the procedure described below entering the RH classifier. Each ensemble setup (i.e. Committee, RH~v1, and RH~v2) has 10 neural network estimators. The architectures of the utilized ML methods are schematically illustrated in Figure~\ref{fig:network}.
    
    The random permutations of features are chosen by first computing the $\chi$-squared and mutual information gain statistics between the features and the SEP presence to assign scores to each feature based on how significant each is in determining whether or not a given flare caused a SEP event. Each feature's score is calculated using this formula:
    \begin{eqnarray}
        s_{\rm i} & = & \sqrt{\chi_{\rm i}^2} + \kappa_{\rm i}
    \end{eqnarray}
    
    Here, $s_{\rm i}$ is a given feature's total score, $\chi_{\rm i}^2$ is the $\chi$-squared statistic between a given feature and SEP presence, and $\kappa_{\rm i}$ is the mutual information gain statistic between a given feature and SEP presence. After each feature is scored based on this formula, the scores are normalized so that the sum of these scores (``feature weights'') equals one. The feature weights are then used as the probabilities that given features with their respective weights will be chosen for a given RH estimator.
    
    Each neural network, including CoNNs, Committee estimators, and RH estimators, has an input layer equal to the number of features being tested by the estimator, a dense layer with an input and output shape of 10, a dropout layer with a probability of 0.2, and an output layer with an output shape equal to the number of predicted features. The networks are implemented using the Skorch library \citep{skorch2017}. The numbers of epochs and learning rates for all CoNN and Committee setups are $n_{\rm epochs}=500$ and $\alpha{}=0.001$ following the hyperparameter optimization discussed later. In addition, RH boosts its epoch counts and learning rates using these formulae:
    \begin{eqnarray}
        n_{\rm epochs}  & = & 500\times{}(2 - x) \\
        \alpha_{\rm RH~v1} & = & 0.001\times{}\left[1 + \ln\left(\dfrac{e^x - x}{\eta} + 1\right)\right] \\
        \alpha_{\rm RH~v2} & = & 0.001\times{}(0.5 + x)
        \label{eq:RHsetting}
    \end{eqnarray}
    	
    Here, $x$ is the total sum of feature weights for a given estimator, $n_{\rm epochs}$ is the number of epochs during the training process, $\alpha$ is the learning rate, and $e$ is the base of the natural logarithm. The parameter $\eta{}=\frac{e^x\Sigma{}x}{n_{\rm features}}$, where $n_{\rm features}$ is the number of features selected for the given estimator, and $\Sigma{}x$ is the total weight of all features in all estimators within the ensemble. The RH classifier is tested twice, with $\alpha$ equal to $\alpha_{1}$ in the first test and $\alpha_{2}$ in the second. Outcomes are predicted by putting prediction data through each of the estimators constructed during the training phase and seeing what each estimator chooses as a predicted result. Each committee considers all results by all estimators as equal, using a simple plurality vote to determine which class a given datum belongs to. Each RH considers each estimator’s value in a classification vote as equal to the sum of the feature weights said estimator’s input features have. For all neural networks, including CoNNs, Committee estimators, and RH estimators, the Adam optimizer \citep{Kingma2015} is used, the cross-entropy loss function is used with balanced class weights, and overfit prevention measures including dropout layers with probabilities of 0.2 and data shuffling are used.
    
    Let us consider an example of feature weights in more detail. If a given flare’s SXR peak flux had a feature weight of $0.25$, its emission measure peak value had one of $0.1$, its temperature peak value had one of $0.05$, and its duration had one of $0.01$, and the total sum of all the feature weights was $1$, the peak SXR flux would have a probability of $0.25$ of being chosen to be in an RH estimator, the emission measure peak flux would have one of $0.1$, etc. A CoNN and a Committee, however, would consider all available features equally as input features. During training, an RH estimator that uses all four of these parameters would go through 15 epochs with learning rates of approximately $\alpha_{\rm RH~v1}=0.00162$ in the first test and $\alpha_{\rm RH~v2}=0.00186$ in the second. A CoNN and a Committee estimator in this example, however, would each only go through 10 epochs with a learning rate of $\alpha=0.001$, since they cannot automatically calculate these parameters based on feature selection. When deciding, each RH estimator would use the sums of its feature weights as values, so an estimator with these four parameters would have a value of 0.5 when voting. Each Committee estimator would have a value of 1, since, again, no mechanism exists to determine how to calculate these figures based on feature selection.

    Tuning of hyperparameters is of known importance for machine learning. For the neural network-based approaches, hyperparameters may include those related to the network architecture (number of hidden layers, neurons in each layer, activation functions, etc.) and the training process (number of epochs, learning rate, optimizer, and regularization parameters, etc.). The parameter space increases even further if considering ensemble approaches. Exploring the entire parameter space is very costly. Therefore, in this work, we restrict the model architectures to those illustrated in Figure~\ref{fig:network}, leaving only the training process-related parameters for optimization. We also notice that the ensemble approaches used in this work have the common CoNN structure as their basis; therefore, optimizing the training for CoNN should deliver the optimal training for the ensemble approaches as well. We explore a grid of the learning rates of $\alpha{}=\{0.01,0.0025,0.001,0.0005\}$, and epoch counts up to $n_{\rm epochs}=2000$, with 10 experiments per each parameter pair, to find the optimal, yet not relatively costly, CoNN training procedure.

\begin{figure}[ht]
        \centering
        \includegraphics[width=1.0\linewidth]{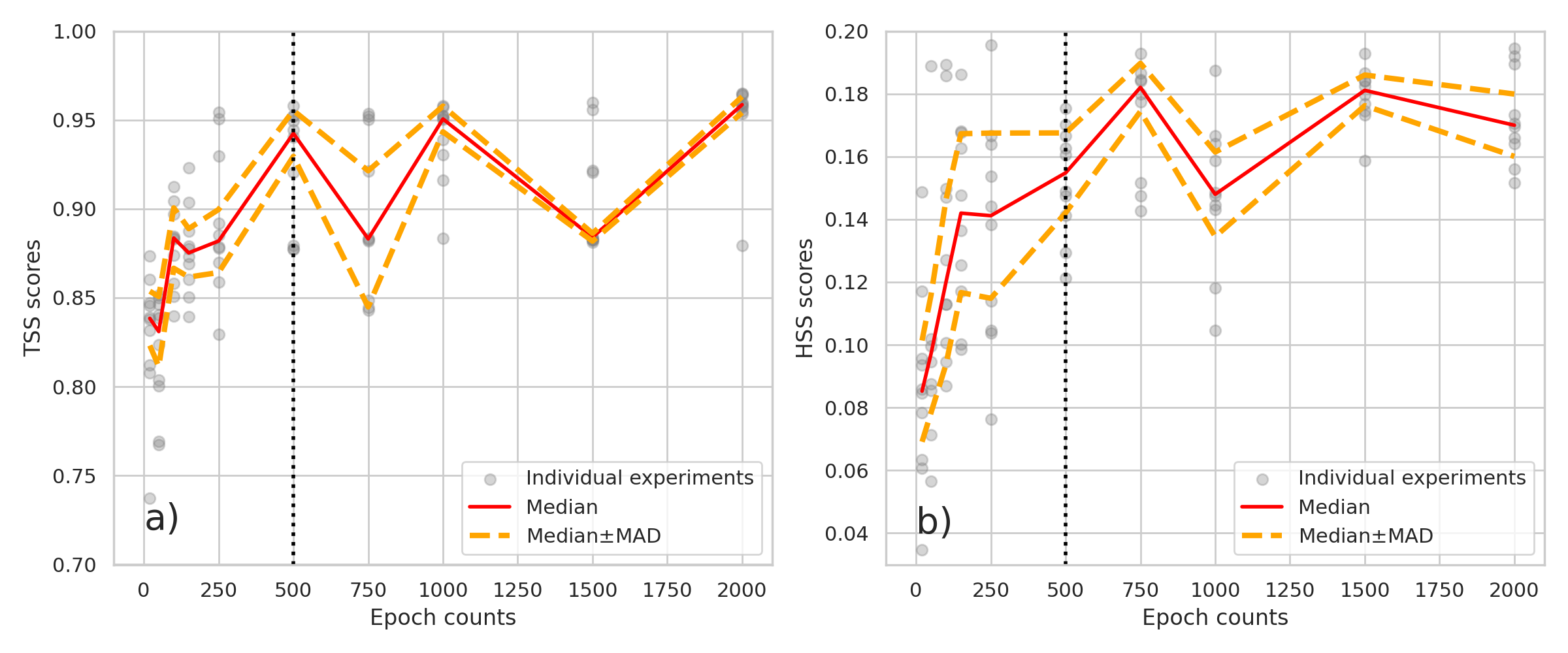}
        \caption{Median (red) and median absolute deviation (orange) performance of the CoNN for the different number of the training epochs and the learning rate of $\alpha{}=0.001$. Panel (a) corresponds to the TSS score, and panel (b) corresponds to the HSS score. Gray points indicate the results of individual experiments. black dotted line indicates the number of epochs considered as optimal and is used for the study.}
        \label{fig:optimization}
    \end{figure}

    Our results for optimization of CoNN with the $\alpha{}=0.001$ are illustrated in Figure~\ref{fig:optimization}. We note here that the further decrease of the learning rate to $\alpha{}=0.0005$ did not lead to an increase in the HSS and TSS scores. As one can see, the CoNN performance roughly increases with the increase in the number of epochs. We decided to select $n_{\rm epochs}=500$ for the prediction experiments: as evident, at this number of epochs trained, the classifier's TSS score starts to decrease significantly for the first time. It is also a relatively low number of epoch counts which provides a possibility to perform a massive number of runs for CoNN and ensemble approaches while evaluating their performance. However, we note here that further fine-tuning of the hyperparameters is possible for this problem.

    Overall, the hyperparameters of the RH algorithm include the architectures of the individual estimators, the implementation of the particular weighting scheme, feature importance estimation, the number of features to use in each estimator, and the adjustment of the RH training parameters, etc. The RH implementations considered here were only constructed to demonstrate a proof-of-concept for RH. In particular, we show further that RH~v2 is performing better compared to RH~v1 on the problem of SEP prediction. This illustrates that different realizations of the RH algorithm may work differently on the same data set and that a careful tuning of RH parameters and options is required to construct the most optimal model for the particular task.
    
    The CoNN and each ensemble member have two output channels predicting the likelihood of the entry belonging to the positive (SEP-associated) or the negative (non-SEP) class. For the binary classification, the class of the larger likelihood is chosen as the prediction. To issue the ensemble prediction, an RH classifier chooses the result that receives the highest number of weighted votes. We demonstrate the individual elements of the confusion matrix (true positive predictions, $TP$, true negative predictions, $TN$, false positive predictions, $FP$, and false negative predictions, $FN$) for each approach averaged over 50 random train-test splits. We also use various metrics to compare binary classification predictions including accuracy, true skill score (TSS), Heidke skill score 2 (HSS), precision, and recall. For a definition of these metrics, see, for example, \citet{bobra2015} and references therein. TSS and HSS scores are also discussed in more detail in Section~\ref{sec:results}. In order to mitigate the susceptibility of the accuracy metric to the high-class imbalance of the data set, we consider in addition the balanced accuracy, defined here as:
    \begin{eqnarray}
        BA = \dfrac{1}{2}\left( \dfrac{TP}{TP+FN} + \dfrac{TN}{TN+FP} \right).
        \label{eq:BalancedAccuracyg}
    \end{eqnarray}

    We also construct the Receiver Operating Characteristic curves for each classifier and calculate the area under the curve (ROC\_AUC) as a prediction metric. The probabilities of classes for the individual estimators are obtained from likelihoods using a softmax function. The positive class probabilities (which are to construct the ROCs) for the RH tests are calculated by taking weighted averages of the probability predictions of each estimator, with each weight being the total sum weight for each estimator following the normalizations of all weight sums.

\section{Results and Discussion}
    \label{sec:results}

\begin{figure}[ht]
        \centering
        \includegraphics[width=0.49\linewidth]{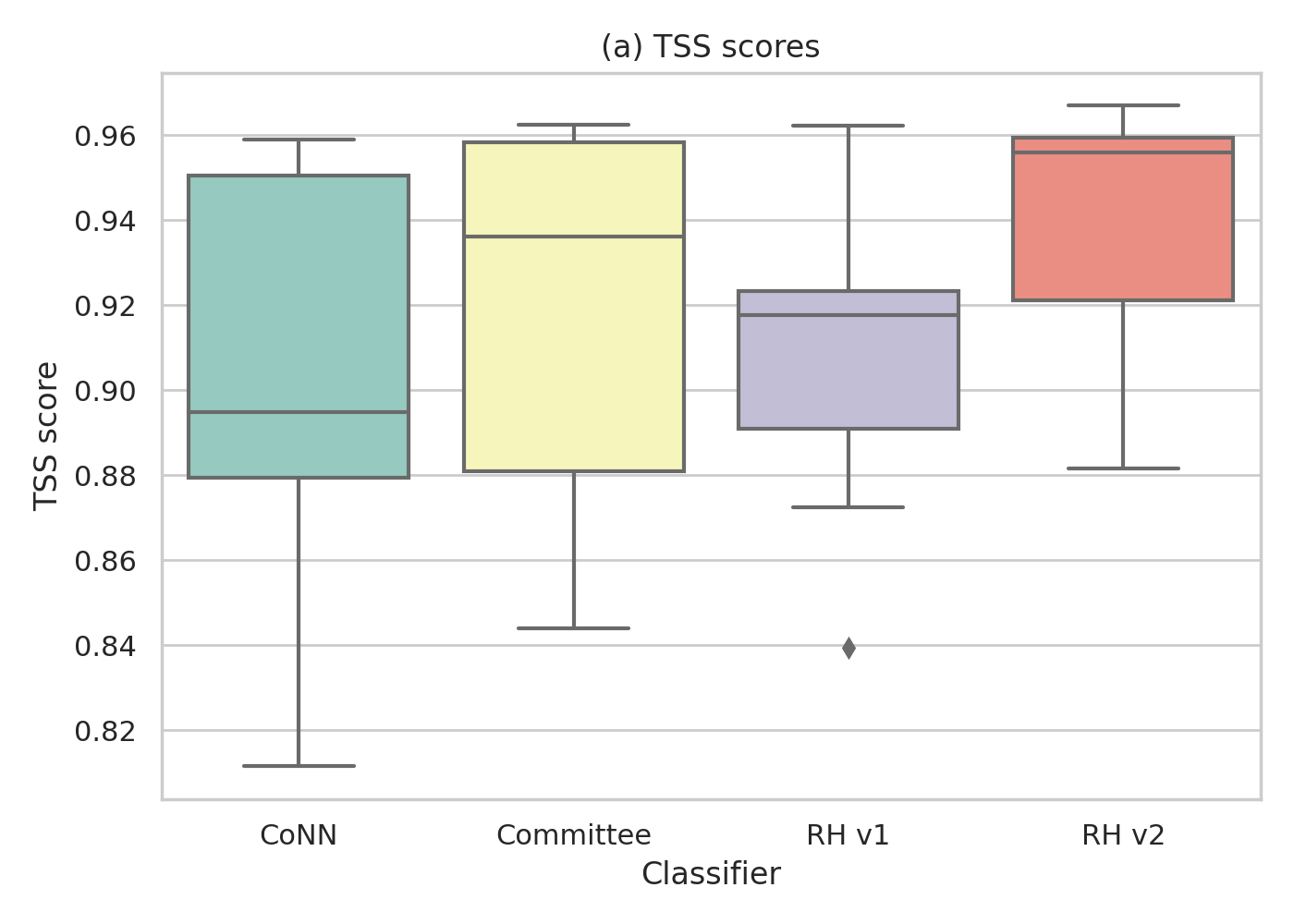}
        \includegraphics[width=0.49\linewidth]{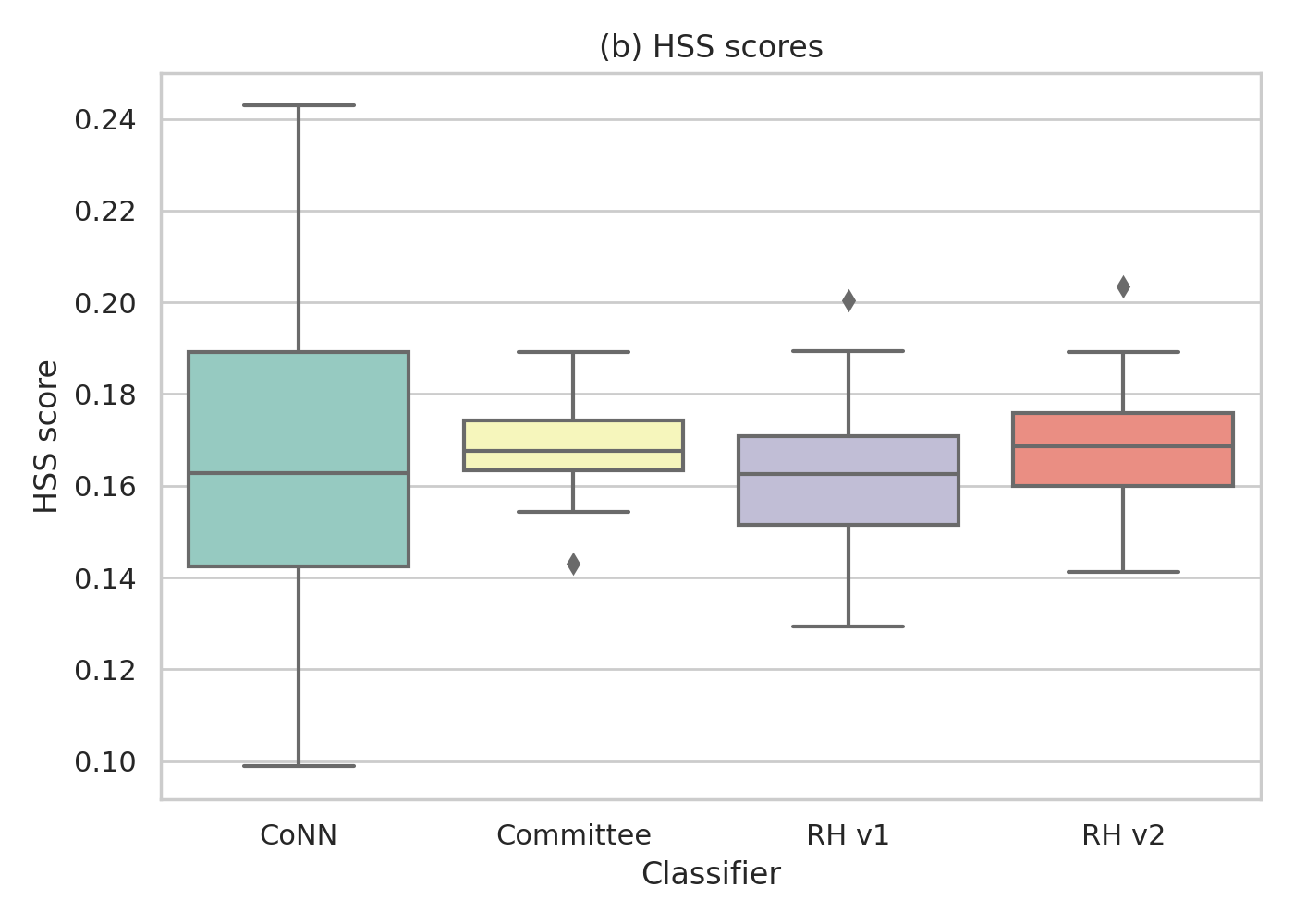}
        \caption{Box-and-whiskers plots summarizing the performances of the CoNN, Committee, and two considered versions of RH, over the 50 experiments in terms of the TSS and HSS scores. Each colored rectangle spans through the second and third quartiles of the scores, with the horizontal bar marking the median. The whiskers indicate the locations of the last individual test results within the interquartile range from the boxes. The rhombus points mark outliers outside the interquartile range from the boxes.}
        \label{fig:boxplots}
    \end{figure}

\begin{figure}[ht]
        \centering
        \includegraphics[width=1.00\linewidth]{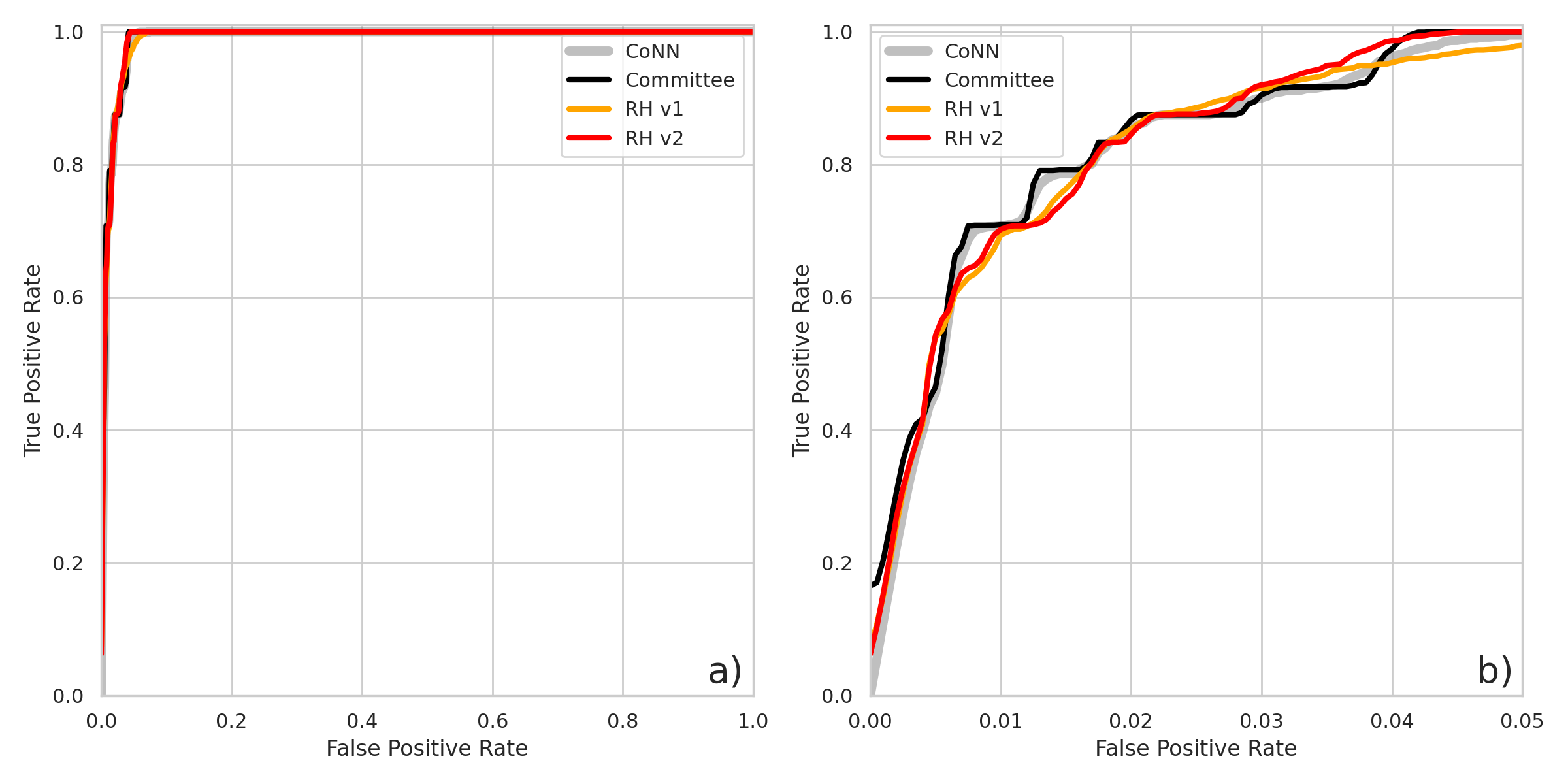}
        \caption{(a) Receiver Operating Characteristic (ROC) curves averaged over 50 experiments for the classifiers considered in the current work. (b) The same curves zoomed in for the small False Positive Rates only.}
        \label{fig:ROCcurves}
    \end{figure}

    The results of the classification algorithms employed in this study in terms of confusion matrix elements and various prediction scores are presented in Table~\ref{tab:averages} (summary results for all classifiers as the average scores and standard deviations) and Table~\ref{tab:medians} (summary results for all classifiers as the median scores and median absolute deviations). The statistics of the results are also illustrated in the box-and-whisker plot presented in Figure~\ref{fig:boxplots} and the Receiver Operating Characteristic (ROC) curves averaged over 50 experiments in Figure~\ref{fig:ROCcurves}. One of the indicators of the sufficiency of 50 experiments for the statistical robustness of the conclusions is that the mean and median values agree with each other to the last significant digit for almost every single metric and classifier considered.
    
    \begin{table}[ht]
        \centering
        \begin{tabular}{| c | c | c | c | c |}
        \hline
        Algorithm / Metrics            &   CoNN    &   Committee    &   RH v1    &   RH v2  \\
        \hline
        TN & 5233.0$\pm$74.4 & 5250.9$\pm$18.9 & 5244.2$\pm$25.5 & 5245.2$\pm$22.8 \\
        \hline
        FP & 237.0$\pm$74.4 & 219.0$\pm$18.9 & 225.7$\pm$25.5 & 224.8$\pm$22.8 \\
        \hline
        FN & 1.2$\pm$1.3 & 0.8$\pm$0.9 & 1.1$\pm$0.7 & 0.4$\pm$0.6 \\
        \hline
        TP & 22.8$\pm$1.3 & 23.2$\pm$0.9 & 22.9$\pm$0.7 & 23.6$\pm$0.6 \\
        \hline
        Precision & 0.094$\pm$0.022 & 0.096$\pm$0.005 & 0.093$\pm$0.009 & 0.096$\pm$0.008 \\
        \hline
        Recall & 0.945$\pm$0.053 & 0.966$\pm$0.038 & 0.956$\pm$0.031 & 0.985$\pm$0.025 \\
        \hline
        Accuracy & 0.957$\pm$0.013 & 0.960$\pm$0.003 & 0.959$\pm$0.005 & 0.959$\pm$0.004 \\
        \hline
        Balanced Accuracy & 0.953$\pm$0.021 & 0.963$\pm$0.018 & 0.957$\pm$0.014 & 0.972$\pm$0.011 \\
        \hline
        TSS & 0.906$\pm$0.043 & 0.926$\pm$0.035 & 0.915$\pm$0.029 & 0.944$\pm$0.023 \\
        \hline
        HSS & 0.163$\pm$0.036 & 0.168$\pm$0.009 & 0.163$\pm$0.014 & 0.168$\pm$0.013 \\
        \hline
        ROC\_AUC & 0.9903$\pm$0.0005 & 0.9907$\pm$0.0001 & 0.9901$\pm$0.0005 & 0.9906$\pm$0.0003 \\
        \hline
        \end{tabular}
        \caption{Average values and standard deviations of the performances of the classifiers considered in this paper.}
        \label{tab:averages}
    \end{table}

    \begin{table}[ht]
        \centering
        \begin{tabular}{| c | c | c | c | c |}
        \hline
        Algorithm / Metrics            &   CoNN    &   Committee    &   RH v1    &   RH v2  \\
        \hline
        TN & 5233.0$\pm$46.5 & 5250.9$\pm$10.0 & 5244.2$\pm$20.0 & 5245.2$\pm$15.5 \\
        \hline
        FP & 237.0$\pm$46.5 & 219.0$\pm$10.0 & 225.7$\pm$20.0 & 224.8$\pm$15.5 \\
        \hline
        FN & 1.2$\pm$1.5 & 0.8$\pm$0.5 & 1.1$\pm$0.0 & 0.4$\pm$0.0 \\
        \hline
        TP & 22.8$\pm$1.5 & 23.2$\pm$0.5 & 22.9$\pm$0.0 & 23.6$\pm$0.0 \\
        \hline
        Precision & 0.094$\pm$0.016 & 0.096$\pm$0.003 & 0.093$\pm$0.006 & 0.096$\pm$0.005 \\
        \hline
        Recall & 0.949$\pm$0.063 & 0.966$\pm$0.021 & 0.956$\pm$0.0 & 0.985$\pm$0.0 \\
        \hline
        Accuracy & 0.957$\pm$0.008 & 0.960$\pm$0.002 & 0.959$\pm$0.004 & 0.959$\pm$0.003 \\
        \hline
        Balanced Accuracy & 0.953$\pm$0.021 & 0.963$\pm$0.011 & 0.957$\pm$0.005 & 0.972$\pm$0.002 \\
        \hline
        TSS & 0.906$\pm$0.042 & 0.926$\pm$0.023 & 0.915$\pm$0.010 & 0.944$\pm$0.005 \\
        \hline
        HSS & 0.163$\pm$0.026 & 0.168$\pm$0.005 & 0.163$\pm$0.010 & 0.168$\pm$0.008 \\
        \hline
        ROC\_AUC & 0.9903$\pm$0.0002 & 0.9907$\pm$0.0001 & 0.9901$\pm$0.0003 & 0.9906$\pm$0.0001 \\
        \hline
        \end{tabular}
        \caption{Median values and median absolute deviations (computed as median values of the absolute deviations of the individual scores from the median) of the performances of the classifiers considered in this paper.}
        \label{tab:medians}
    \end{table}

    Tables~\ref{tab:averages}~and~\ref{tab:medians} indicate that the ensemble approaches are performing better, in general, than the CoNN classifiers with respect to the measures typically used in space weather forecasting, $HSS$ and $TSS$, both in terms of the mean and median values. For example, the $TSS$ score had its median value of $TSS$=0.906$\pm$0.042 for the CoNN classifier and increased to $TSS$=0.926$\pm$0.023, $TSS$=0.915$\pm$0.010, and $TSS$=0.944$\pm$0.005 (RH~v2) for the Committee and two RH ensemble classifiers. Although the $HSS$ scores were relatively low, they still demonstrated either no drop or an increase from $HSS=$0.163$\pm$0.026 to $HSS=$0.168$\pm$0.005, $HSS=$0.163$\pm$0.010, and $HSS=$0.168$\pm$0.008 when transitioning from the CoNN to ensembles. This demonstrates that the ensemble approaches are performing better, on average, with respect to the CoNN classifier. On the other hand, we point out here that both the TSS and HSS scores for all classifiers almost always intersected within the uncertainties (either the standard deviation or the median absolution deviation) arising from the results of individual experiments. The overall closeness of the performance of the classifiers is evident as well from the ROC curves presented in Figure~\ref{fig:ROCcurves} which experience a significant overlap. The good performance of the ensemble classifiers was previously noticed in the works of \citet{Aminalragia2021} for the Committee approach and \citet{Lavasa2021} for the random forest classifier. Interestingly, the case performances of the CoNN classifier may even outperform the individual ensemble classifier tests (as evident from the upper boundary for interquartile ranges for CoNN in Figure~\ref{fig:boxplots}), which reveals the importance of evaluation of the methods on several train-test splits and demonstration of its robustness with respect to the random splitting.

    Another pattern evident from Tables~\ref{tab:averages}~and~\ref{tab:medians} is the noticeable differences between the standard deviations or median absolute deviations for the CoNN and ensemble classifiers. For example, the mean $TSS$ score and its standard deviation for CoNN is $TSS$=0.906$\pm$0.043 compared to the $TSS$=0.926$\pm$0.035 (Committee), $TSS$=0.915$\pm$0.029 (RH~v1), and $TSS$=0.944$\pm$0.023 (RH~v2). The tendency is even sharper for the medians and median absolute deviations (compare $TSS$=0.906$\pm$0.042 for CoNN with $TSS$=0.926$\pm$0.023, $TSS$=0.915$\pm$0.010, and $TSS$=0.944$\pm$0.005 for aforementioned ensemble approaches). The tendency of lower spread remains the same for all other scores considered in the tables. Figure~\ref{fig:boxplots} also indicates that the interquartile range and the span of whiskers (indicating the spread of individual experiments outside of the interquartile range) is typically larger for CoNN, especially in the case of $HSS$ skill score. Overall, such behavior indicates the relative robustness of the ensemble approaches with respect to the random train-test splits for the data set and the training process, while the training of the individual classifiers may fail. Therefore, the increase in the complexity of these ensemble algorithms is justified by their robust performance on the imbalanced data sets \citep{galar2012}.
    
    Tables~\ref{tab:averages}~and~\ref{tab:medians} also demonstrate that the RH classifier (1) does not necessarily outperform the Committee approach, and (2) depends on the selection of its parameters. This is concluded from the fact that the performance of the RH~v2 is, on average, the same or better, with respect to almost any metrics than that of the Committee or RH~v1. Another noticeable difference between the Committee and RH classifiers is the step-wise behavior of the ROC curve for the Committee compared to a smoother curve for the RH, evident in Figure~\ref{fig:ROCcurves}. The key difference between the RH and the Committee classifier is in the selection of features used for each individual ensemble member. While the committee members use all features available, RH members use the down-selected number of features (either 4 or 6 out of 12 in our case) and use the deterministic algorithm of the contribution of each Committee member to the final result. We may assume that the down-selection of features for each member increases the forecasting scores because it helps to filter out the attention of the individual committee member to the noisy or irrelevant features and prevents its members from ``agreeing'' on identical sets of model parameters (weights and biases). This confirms the importance of the feature selection process, which remains an active topic in predicting solar transient events \citep{bobra2015,sadykov2017b,Yeolekar2021}. Also, although the Committee approach \citep{Aminalragia2021} helps to reduce the ``reliance on chance'' in terms of the convergence of the network parameters (weights and biases) to the local or global minima, it still contains similarly-structured CoNNs as ensemble members. The RH introduces a more diverse population of ensemble members with the variable down-selected set of features as an input, which can be more beneficial than having full-scale but nearly identical learners.

    As noted earlier, the results in Tables~\ref{tab:averages}~and~\ref{tab:medians} indicate relatively low values for the precision and HSS scores for all classifiers tested, including the RH classifiers. At the same time, the corresponding TSS scores are high. To understand the reason behind this behavior of the models, let us indicate some patterns in our SEP prediction. Looking at the median confusion matrix elements, one can notice the RH classifiers that, arranged by larger to smaller, $TN\approx{}5244\gg{}FP\approx{}226\gg{}TP\approx{}22.9\gg{}FN\approx{}1.1$ and that $TN\approx{}5245\gg{}FP\approx{}225\gg{}TP\approx{}23.6\gg{}FN\approx{}0.4$. Assuming that one can neglect the term of the next order of smallness, one can rewrite the metrics of interest as:

    \begin{eqnarray}
        &Precision = \dfrac{TP}{TP+FP} \approx \dfrac{TP}{FP} \\
        &HSS = \dfrac{2\cdot{}(TP\cdot{}TN-FP\cdot{}FN)}{(TP+FN)(FN+TN)+(TP+FP)(FP+TN)} \approx \nonumber \\
        &\approx{}\dfrac{2\cdot{}TP\cdot{}TN}{TP\cdot{}TN + FP\cdot{}TN} \approx \dfrac{2\cdot{}TP}{TP+FP} = 2\cdot{}Precision \\
        &TSS = \dfrac{TP}{TP+FN} - \dfrac{FP}{FP+TN} = 1 - \dfrac{FN}{TP+FN} - \dfrac{FP}{FP+TN} \approx 1 - \dfrac{FN}{TP} - \dfrac{FP}{TN} \rightarrow 1
    \end{eqnarray}

    Both the precision and HSS scores, under the conditions for the confusion matrix elements indicated above, are determined by the $TP/FP$ ratio, which is of the order of $\sim$1/10 for the models implemented in this study. At the same time, the TSS score is very close to 1 because both subtrahends are small: $FN/TP\sim$1/20 or smaller, and $FP/TN\sim$1/23. While capturing almost every single SEP flare in the test data set (which is 24 events on average), the models produce almost ten times larger number of false alarms than the number of SEP flares (with the median values of $FP=225.7\pm{}20.0$ and $FP=224.8\pm{}15.5$ for RH~v1 and RH~v2, correspondingly). The considered data set is also highly imbalanced given that the ratio of the positive to negative samples is $\sim$1/285. The conditions above limit the HSS score to $HSS\approx{}2\times{}TP/FP{}\leq{}2\times{}P/FP$. They also make the HSS scores to be very susceptible to the change in the number of false alarms. For example, if one manages to decrease the number of false alarms twice ($FP=224.8\rightarrow{}FP=112.4$ and $TN=5245.2\rightarrow{}TN=5357.6$ for RH~v2), the corresponding HSS score would increase almost twice, to $HSS\approx{}0.290$. At the same time, the TSS score would increase from $TSS\approx{}0.944$ to $TSS\approx{}0.963$, experiencing a mild relative change. One can conclude that for highly imbalanced data sets the small HSS scores are related to the large number of false alarms produced by the model (with respect to the number of positive samples in the test data set) and may still be accompanied by TSS scores very close to one.
    
    Also, we note that while the Heidke Skill Score 2 (HSS) is often annotated as a measure of the performance with respect to a random chance forecast, the forecast presented here is definitely far from being random: with RH~v1, the missed event rate $FN/(FN+TP)\approx{}0.046$ is low (almost every SEP event is hit for each trial) and the false alarm rate $FP/(FP+TN)\approx{}0.041$ is low as well, while with RH~v2, the missed event rate is even more promising $FN/(FN+TP)\approx{}0.017$ and the false alarm rate is at the still the same value of approximately $0.041$. Nevertheless, the HSS scores are not so strongly deviating from $0$ (on average, $HSS=0.163\pm{}0.014$ with RH~v1 and $HSS=0.168\pm{}0.013$ with RH~v2). Therefore, we argue that it is not correct to associate low $HSS$ scores with the forecast being close to a random chance forecast. Moreover, the example in the previous paragraph demonstrates that it is very beneficial to consider the $HSS$ scores complementary to the $TSS$ scores for forecasting purposes. The $HSS$ score is much more sensitive to the decrease of the false alarms, $FP$, if compared to the $TSS$. Therefore, it would allow one to better differentiate between the models with approximately the same number of missed events based on the number of false alarms they produce.

    Although not tested for the all-clear forecasting explicitly, the classification approaches implemented here demonstrate usefulness with respect to the all-clear setting. We note here that the all-clear SEP forecasting is typically defined as the whole-disk endeavor which also has to be accompanied by the specification of the time window for which the forecast is issued. Here we discuss a related, but not identical, problem when one would like to predict very reliably every SEP-active flare (with the missed events being very undesirable) yet maintaining a low false alarm rate. Here the very low rate of missed events (the average rate of the missed events is $FN/(FN+TP)\approx{}0.05$ even for CoNN) is what is typically desirable for such a type of forecast \citep{sadykov2021}. For the RH~v2 classifier, the virtually zero rate of missed events ($FN/(FN+TP)\approx{}0.017$) renders the RH viable for this task. Although the median number of the $FN$ was still relatively low for the CoNN classifier ($FN=1.2\pm{}1.5$, see Table~\ref{tab:medians}) and the Committee approach ($FN=0.8\pm{}0.5$), the RH~v2 has even lower $FN=0.4\pm{}0.0$. The median absolute deviation of 0.0 indicates that more than half (i.e., more than 25 out of 50) trials of RH~v2 had zero false negative predictions as an outcome. Overall, we can also potentially expect, that certain configurations of the RH approach can be useful in delivering near-zero missed event forecasting with low false-alarm rates for other machine-learning problems, including those involving highly imbalanced classes.

\section{Conclusion}
    \label{sec:discussion}

    In this work, we have introduced an ensemble algorithm~--- a Random Hivemind (RH)~--- and compared two selected implementations of it with respect to the Conventional Neural Network (CoNN) and a Committee ensemble approach for CoNNs. The comparison was done for the problem of the prediction of Solar Energetic Particle (SEP) events based on the properties of the host soft X-ray flares. The key outcomes of our work are as follows:

    \begin{itemize}
        \item The performance of the RH algorithm depends on the implementation and training parameters (which may include the number of input features, their weighting schemes, learning rate boosting strategies, etc.). The corresponding TSS and HSS scores for the RH~v1 and RH~v2 implementations are $TSS=0.915\pm{}0.010$ and $TSS=0.944\pm{}0.005$, and $HSS=0.163\pm{}0.010$ and $HSS=0.168\pm{}0.008$, correspondingly.
        \item Both ensemble approaches (Committee and RH) demonstrate the robustness of their performance with respect to the random train-test splits for the data set, which was reflected in the low standard deviations or median absolute deviations. Although often performing comparably to the committee approach in terms of the forecasting metrics, CoNN demonstrated much higher standard deviations (and often higher median absolute deviations).
        \item Both ensemble approaches demonstrated similar or better performance in terms of mean and median values compared to the metrics typically used in space weather forecasting, $HSS$ and $TSS$. One can compare the median $TSS=0.906\pm{}0.042$ for CoNN with $TSS=0.926\pm{}0.023$, $TSS=0.915\pm{}0.010$, and $TSS=0.944\pm{}0.005$ for the committee, RH~v1, and RH~v2 correspondingly, and $HSS=0.163\pm{}0.026$ with $HSS=0.168\pm{}0.005$, $HSS=0.163\pm{}0.010$, and $HSS=0.168\pm{}0.008$.
        \item The RH~v2 ensemble classifier performs better, on average, than the Committee, CoNN, and RH~v1 approaches in terms of almost every metric and delivers consistent results over the ten random train-test split experiments.
        \item The performance of all classifiers, including RH, demonstrated relatively low precision and $HSS$ scores for the SEP prediction problem. Nevertheless, it is very beneficial to consider the $HSS$ as a complementary metric for the forecast as it is more susceptible to a decrease in the false alarm rate than the $TSS$. This leads to a better differentiation between the models with approximately the same number of missed events based on the number of false alarms they produce.
        \item All classifiers had a very low number of false negative predictions. Median values of $FN=1.2\pm{}1.5$, $FN=0.8\pm{}0.5$, $FN=1.1\pm{}0.0$, and $FN=0.4\pm0.0$,  were measured for the CoNN, Committee, RH~v1, and RH~v2 classifiers, respectively. However, the robustness of the RH classifiers noted previously, especially for the case of RH~v2, makes it the viable candidate for employment in solving the ``all-clear''-like forecasting problem for SEP-active flares.
    \end{itemize}

    From the results above, we can conclude that RH is a valid machine learning algorithm that can perform well despite class imbalance. RH is performing, on average, comparably or better to CoNNs and unweighted, identical CoNN committee machines. Further studies of the RH approach (including different implementations for the feature weights and handling, learning rate and epoch number adjustments, and the flare class boundaries considered for RH training) are required to understand its potential in general and specifically for space weather prediction purposes, including ``all-clear'' forecasting of SEPs.

\section*{Acknowledgments} The authors thank the anonymous referees for the suggestions which significantly improved the quality of the manuscript. This research was supported by NASA Early Stage Innovation program grant 80NSSC20K0302, NASA LWS grant 80NSSC19K0068, NSF EarthCube grants 1639683, 1743321, and 1927578, and NSF grant 1835958. VMS acknowledges the NSF FDSS grant 1936361 and NSF grant 1835958.

\bibliographystyle{model1-num-names} 
\bibliography{refs}

\begin{thebibliography}{38}
\expandafter\ifx\csname natexlab\endcsname\relax\def\natexlab#1{#1}\fi
\providecommand{\url}[1]{\texttt{#1}}
\providecommand{\href}[2]{#2}
\providecommand{\path}[1]{#1}
\providecommand{\DOIprefix}{doi:}
\providecommand{\ArXivprefix}{arXiv:}
\providecommand{\URLprefix}{URL: }
\providecommand{\Pubmedprefix}{pmid:}
\providecommand{\doi}[1]{\href{http://dx.doi.org/#1}{\path{#1}}}
\providecommand{\Pubmed}[1]{\href{pmid:#1}{\path{#1}}}
\providecommand{\bibinfo}[2]{#2}
\ifx\xfnm\relax \def\xfnm[#1]{\unskip,\space#1}\fi
\bibitem[{{Balch}(1999)}]{balch1999}
\bibinfo{author}{C.~C. {Balch}},
\newblock \bibinfo{title}{{SEC proton prediction model: verification and analysis}},
\newblock \bibinfo{journal}{Radiation Measurements} \bibinfo{volume}{30} (\bibinfo{year}{1999}) \bibinfo{pages}{231--250}.
\bibitem[{{Balch}(2008)}]{balch2008}
\bibinfo{author}{C.~C. {Balch}},
\newblock \bibinfo{title}{{Updated verification of the Space Weather Prediction Center's solar energetic particle prediction model}},
\newblock \bibinfo{journal}{Space Weather} \bibinfo{volume}{6} (\bibinfo{year}{2008}) \bibinfo{pages}{S01001}.
\bibitem[{{Bain} et~al.(2021){Bain}, {Steenburgh}, {Onsager}, and {Stitely}}]{bain2021}
\bibinfo{author}{H.~M. {Bain}}, \bibinfo{author}{R.~A. {Steenburgh}}, \bibinfo{author}{T.~G. {Onsager}}, \bibinfo{author}{E.~M. {Stitely}},
\newblock \bibinfo{title}{{A Summary of National Oceanic and Atmospheric Administration Space Weather Prediction Center Proton Event Forecast Performance and Skill}},
\newblock \bibinfo{journal}{Space Weather} \bibinfo{volume}{19} (\bibinfo{year}{2021}) \bibinfo{pages}{e2020SW002670}.
\bibitem[{{Reames}(2021)}]{reames2021}
\bibinfo{author}{D.~V. {Reames}}, \bibinfo{title}{{Solar Energetic Particles. A Modern Primer on Understanding Sources, Acceleration and Propagation}}, volume \bibinfo{volume}{978}, \bibinfo{year}{2021}. \DOIprefix\doi{10.1007/978-3-030-66402-2}.
\bibitem[{{Ryan} et~al.(2012){Ryan}, {Milligan}, {Gallagher}, {Dennis}, {Tolbert}, {Schwartz}, and {Young}}]{ryan2012}
\bibinfo{author}{D.~F. {Ryan}}, \bibinfo{author}{R.~O. {Milligan}}, \bibinfo{author}{P.~T. {Gallagher}}, \bibinfo{author}{B.~R. {Dennis}}, \bibinfo{author}{A.~K. {Tolbert}}, \bibinfo{author}{R.~A. {Schwartz}}, \bibinfo{author}{C.~A. {Young}},
\newblock \bibinfo{title}{{The Thermal Properties of Solar Flares over Three Solar Cycles Using GOES X-Ray Observations}},
\newblock \bibinfo{journal}{The Astrophysical Journal Supplement Series} \bibinfo{volume}{202} (\bibinfo{year}{2012}) \bibinfo{pages}{11}.
\bibitem[{{Sadykov} et~al.(2019){Sadykov}, {Kosovichev}, {Kitiashvili}, and {Frolov}}]{sadykov2019}
\bibinfo{author}{V.~M. {Sadykov}}, \bibinfo{author}{A.~G. {Kosovichev}}, \bibinfo{author}{I.~N. {Kitiashvili}}, \bibinfo{author}{A.~{Frolov}},
\newblock \bibinfo{title}{{Statistical Properties of Soft X-Ray Emission of Solar Flares}},
\newblock \bibinfo{journal}{The Astrophysical Journal} \bibinfo{volume}{874} (\bibinfo{year}{2019}) \bibinfo{pages}{19}.
\bibitem[{{Garcia}(1994)}]{garcia1994}
\bibinfo{author}{H.~A. {Garcia}},
\newblock \bibinfo{title}{{Temperature and Hard X-Ray Signatures for Energetic Proton Events}},
\newblock \bibinfo{journal}{The Astrophysical Journal} \bibinfo{volume}{420} (\bibinfo{year}{1994}) \bibinfo{pages}{422}.
\bibitem[{{Garcia}(2004)}]{garcia2004}
\bibinfo{author}{H.~A. {Garcia}},
\newblock \bibinfo{title}{{Forecasting methods for occurrence and magnitude of proton storms with solar soft X rays}},
\newblock \bibinfo{journal}{Space Weather} \bibinfo{volume}{2} (\bibinfo{year}{2004}) \bibinfo{pages}{S02002}.
\bibitem[{{Kahler} and {Ling}(2018)}]{kahler2018}
\bibinfo{author}{S.~W. {Kahler}}, \bibinfo{author}{A.~G. {Ling}},
\newblock \bibinfo{title}{{Forecasting Solar Energetic Particle (SEP) events with Flare X-ray peak ratios}},
\newblock \bibinfo{journal}{Journal of Space Weather and Space Climate} \bibinfo{volume}{8} (\bibinfo{year}{2018}) \bibinfo{pages}{A47}.
\bibitem[{{Ling} and {Kahler}(2020)}]{ling2020}
\bibinfo{author}{A.~G. {Ling}}, \bibinfo{author}{S.~W. {Kahler}},
\newblock \bibinfo{title}{{Peak Temperatures of Large Solar X-Ray Flares and Associated CME Speeds and Widths}},
\newblock \bibinfo{journal}{The Astrophysical Journal} \bibinfo{volume}{891} (\bibinfo{year}{2020}) \bibinfo{pages}{54}.
\bibitem[{{Kahler} and {Ling}(2022)}]{kahler2022}
\bibinfo{author}{S.~W. {Kahler}}, \bibinfo{author}{A.~G. {Ling}},
\newblock \bibinfo{title}{{A Comparison of Solar X-Ray Flare Timescales and Peak Temperatures with Associated Coronal Mass Ejections}},
\newblock \bibinfo{journal}{The Astrophysical Journal} \bibinfo{volume}{934} (\bibinfo{year}{2022}) \bibinfo{pages}{175}.
\bibitem[{{Kahler} and {Vourlidas}(2013)}]{kahler2013}
\bibinfo{author}{S.~W. {Kahler}}, \bibinfo{author}{A.~{Vourlidas}},
\newblock \bibinfo{title}{{A Comparison of the Intensities and Energies of Gradual Solar Energetic Particle Events with the Dynamical Properties of Associated Coronal Mass Ejections}},
\newblock \bibinfo{journal}{The Astrophysical Journal} \bibinfo{volume}{769} (\bibinfo{year}{2013}) \bibinfo{pages}{143}.
\bibitem[{{Richardson} et~al.(2018){Richardson}, {Mays}, and {Thompson}}]{richardson2018}
\bibinfo{author}{I.~G. {Richardson}}, \bibinfo{author}{M.~L. {Mays}}, \bibinfo{author}{B.~J. {Thompson}},
\newblock \bibinfo{title}{{Prediction of Solar Energetic Particle Event Peak Proton Intensity Using a Simple Algorithm Based on CME Speed and Direction and Observations of Associated Solar Phenomena}},
\newblock \bibinfo{journal}{Space Weather} \bibinfo{volume}{16} (\bibinfo{year}{2018}) \bibinfo{pages}{1862--1881}.
\bibitem[{{Laurenza} et~al.(2009){Laurenza}, {Cliver}, {Hewitt}, {Storini}, {Ling}, {Balch}, and {Kaiser}}]{Laurenza2009}
\bibinfo{author}{M.~{Laurenza}}, \bibinfo{author}{E.~W. {Cliver}}, \bibinfo{author}{J.~{Hewitt}}, \bibinfo{author}{M.~{Storini}}, \bibinfo{author}{A.~G. {Ling}}, \bibinfo{author}{C.~C. {Balch}}, \bibinfo{author}{M.~L. {Kaiser}},
\newblock \bibinfo{title}{A technique for short-term warning of solar energetic particle events based on flare location, flare size, and evidence of particle escape},
\newblock \bibinfo{journal}{Space Weather Journal} \bibinfo{volume}{7} (\bibinfo{year}{2009}) \bibinfo{pages}{S04008}.
\bibitem[{{Laurenza} et~al.(2018){Laurenza}, {Alberti}, and {Cliver}}]{Laurenza2018}
\bibinfo{author}{M.~{Laurenza}}, \bibinfo{author}{T.~{Alberti}}, \bibinfo{author}{E.~W. {Cliver}},
\newblock \bibinfo{title}{A short-term esperta-based forecast tool for moderate-to-extreme solar proton events},
\newblock \bibinfo{journal}{The Astrophysical Journal} \bibinfo{volume}{857} (\bibinfo{year}{2018}) \bibinfo{pages}{107}.
\bibitem[{{Aminalragia-Giamini} et~al.(2021){Aminalragia-Giamini}, {Raptis}, {Anastasiadis}, {Tsigkanos}, {Sandberg}, {Papaioannou}, {Papadimitriou}, {Jiggens}, {Aran}, and {Daglis}}]{Aminalragia2021}
\bibinfo{author}{S.~{Aminalragia-Giamini}}, \bibinfo{author}{S.~{Raptis}}, \bibinfo{author}{A.~{Anastasiadis}}, \bibinfo{author}{A.~{Tsigkanos}}, \bibinfo{author}{I.~{Sandberg}}, \bibinfo{author}{A.~{Papaioannou}}, \bibinfo{author}{C.~{Papadimitriou}}, \bibinfo{author}{P.~{Jiggens}}, \bibinfo{author}{A.~{Aran}}, \bibinfo{author}{I.~A. {Daglis}},
\newblock \bibinfo{title}{{Solar Energetic Particle Event occurrence prediction using Solar Flare Soft X-ray measurements and Machine Learning}},
\newblock \bibinfo{journal}{Journal of Space Weather and Space Climate} \bibinfo{volume}{11} (\bibinfo{year}{2021}) \bibinfo{pages}{59}.
\bibitem[{Boubrahimi et~al.(2017)Boubrahimi, Aydin, Martens, and Angryk}]{Boubrahimi2017}
\bibinfo{author}{S.~F. Boubrahimi}, \bibinfo{author}{B.~Aydin}, \bibinfo{author}{P.~Martens}, \bibinfo{author}{R.~Angryk},
\newblock \bibinfo{title}{On the prediction of >100 mev solar energetic particle events using goes satellite data},
\newblock in: \bibinfo{booktitle}{2017 IEEE International Conference on Big Data (Big Data)}, \bibinfo{year}{2017}, pp. \bibinfo{pages}{2533--2542}. \DOIprefix\doi{10.1109/BigData.2017.8258212}.
\bibitem[{{Lavasa} et~al.(2021){Lavasa}, {Giannopoulos}, {Papaioannou}, {Anastasiadis}, {Daglis}, {Aran}, {Pacheco}, and {Sanahuja}}]{Lavasa2021}
\bibinfo{author}{E.~{Lavasa}}, \bibinfo{author}{G.~{Giannopoulos}}, \bibinfo{author}{A.~{Papaioannou}}, \bibinfo{author}{A.~{Anastasiadis}}, \bibinfo{author}{I.~A. {Daglis}}, \bibinfo{author}{A.~{Aran}}, \bibinfo{author}{D.~{Pacheco}}, \bibinfo{author}{B.~{Sanahuja}},
\newblock \bibinfo{title}{{Assessing the Predictability of Solar Energetic Particles with the Use of Machine Learning Techniques}},
\newblock \bibinfo{journal}{Solar Physics} \bibinfo{volume}{296} (\bibinfo{year}{2021}) \bibinfo{pages}{107}.
\bibitem[{{Torres} et~al.(2022){Torres}, {Zhao}, {Chan}, and {Zhang}}]{Torres2022}
\bibinfo{author}{J.~{Torres}}, \bibinfo{author}{L.~{Zhao}}, \bibinfo{author}{P.~K. {Chan}}, \bibinfo{author}{M.~{Zhang}},
\newblock \bibinfo{title}{{A Machine Learning Approach to Predicting SEP Events Using Properties of Coronal Mass Ejections}},
\newblock \bibinfo{journal}{Space Weather} \bibinfo{volume}{20} (\bibinfo{year}{2022}) \bibinfo{pages}{e2021SW002797}.
\bibitem[{{Nishizuka} et~al.(2020){Nishizuka}, {Kubo}, {Sugiura}, {Den}, and {Ishii}}]{Nishizuka2020}
\bibinfo{author}{N.~{Nishizuka}}, \bibinfo{author}{Y.~{Kubo}}, \bibinfo{author}{K.~{Sugiura}}, \bibinfo{author}{M.~{Den}}, \bibinfo{author}{M.~{Ishii}},
\newblock \bibinfo{title}{{Reliable Probability Forecast of Solar Flares: Deep Flare Net-Reliable (DeFN-R)}},
\newblock \bibinfo{journal}{The Astrophysical Journal} \bibinfo{volume}{899} (\bibinfo{year}{2020}) \bibinfo{pages}{150}.
\bibitem[{{Afaq} and {Rao}(2020)}]{afaq2020}
\bibinfo{author}{S.~{Afaq}}, \bibinfo{author}{S.~{Rao}},
\newblock \bibinfo{title}{{Significance Of Epochs On Training A Neural Network}},
\newblock \bibinfo{journal}{International Journal of Scientific and Technology Research} \bibinfo{volume}{9} (\bibinfo{year}{2020}) \bibinfo{pages}{485--488}.
\bibitem[{{You} et~al.(2019){You}, {Long}, {Wang}, and {Jordan}}]{you2019}
\bibinfo{author}{K.~{You}}, \bibinfo{author}{M.~{Long}}, \bibinfo{author}{J.~{Wang}}, \bibinfo{author}{M.~I. {Jordan}},
\newblock \bibinfo{title}{{How Does Learning Rate Decay Help Modern Neural Networks?}},
\newblock \bibinfo{journal}{arXiv e-prints}  (\bibinfo{year}{2019}) \bibinfo{pages}{arXiv:1908.01878}.
\bibitem[{{Cunningham}(2000)}]{cunningham2000}
\bibinfo{author}{P.~{Cunningham}},
\newblock \bibinfo{title}{{Overfitting and Diversity in Classification Ensembles based on Feature Selection}},
\newblock \bibinfo{journal}{Trinity's Access to Research Archive}  (\bibinfo{year}{2000}).
\bibitem[{{Ali} et~al.(2023){Ali}, {Sadykov}, {Kosovichev}, {Kitiashvili}, {Oria}, {Nita}, {Illarionov}, {O'Keefe}, {Francis}, {Chong}, {Kosovich}, and {Marroquin}}]{ali2023}
\bibinfo{author}{A.~{Ali}}, \bibinfo{author}{V.~{Sadykov}}, \bibinfo{author}{A.~{Kosovichev}}, \bibinfo{author}{I.~N. {Kitiashvili}}, \bibinfo{author}{V.~{Oria}}, \bibinfo{author}{G.~M. {Nita}}, \bibinfo{author}{E.~{Illarionov}}, \bibinfo{author}{P.~M. {O'Keefe}}, \bibinfo{author}{F.~{Francis}}, \bibinfo{author}{C.-J. {Chong}}, \bibinfo{author}{P.~{Kosovich}}, \bibinfo{author}{R.~D. {Marroquin}},
\newblock \bibinfo{title}{{Predicting Solar Proton Events of Solar Cycles 22-24 using GOES Proton \& Soft X-Ray Flux Statistics}},
\newblock \bibinfo{journal}{arXiv e-prints}  (\bibinfo{year}{2023}) \bibinfo{pages}{arXiv:2303.05446}.
\bibitem[{{Sadykov} et~al.(2021){Sadykov}, {Kosovichev}, {Kitiashvili}, {Oria}, {Nita}, {Illarionov}, {O'Keefe}, {Jiang}, {Fereira}, and {Ali}}]{sadykov2021}
\bibinfo{author}{V.~{Sadykov}}, \bibinfo{author}{A.~{Kosovichev}}, \bibinfo{author}{I.~{Kitiashvili}}, \bibinfo{author}{V.~{Oria}}, \bibinfo{author}{G.~M. {Nita}}, \bibinfo{author}{E.~{Illarionov}}, \bibinfo{author}{P.~{O'Keefe}}, \bibinfo{author}{Y.~{Jiang}}, \bibinfo{author}{S.~{Fereira}}, \bibinfo{author}{A.~{Ali}},
\newblock \bibinfo{title}{{Prediction of Solar Proton Events with Machine Learning: Comparison with Operational Forecasts and ``All-Clear'' Perspectives}},
\newblock \bibinfo{journal}{arXiv e-prints}  (\bibinfo{year}{2021}) \bibinfo{pages}{arXiv:2107.03911}.
\bibitem[{{Stumpo} et~al.(2021){Stumpo}, {Benella}, {Laurenza}, {Alberti}, {Consolini}, and {Marcucci}}]{Stumpo2021}
\bibinfo{author}{M.~{Stumpo}}, \bibinfo{author}{S.~{Benella}}, \bibinfo{author}{M.~{Laurenza}}, \bibinfo{author}{T.~{Alberti}}, \bibinfo{author}{G.~{Consolini}}, \bibinfo{author}{M.~F. {Marcucci}},
\newblock \bibinfo{title}{{Open Issues in Statistical Forecasting of Solar Proton Events: A Machine Learning Perspective}},
\newblock \bibinfo{journal}{Space Weather Journal} \bibinfo{volume}{19} (\bibinfo{year}{2021}) \bibinfo{pages}{e2021SW002794}.
\bibitem[{{Ahmadzadeh} et~al.(2021){Ahmadzadeh}, {Aydin}, {Georgoulis}, {Kempton}, {Mahajan}, and {Angryk}}]{ahmadzadeh2021}
\bibinfo{author}{A.~{Ahmadzadeh}}, \bibinfo{author}{B.~{Aydin}}, \bibinfo{author}{M.~K. {Georgoulis}}, \bibinfo{author}{D.~J. {Kempton}}, \bibinfo{author}{S.~S. {Mahajan}}, \bibinfo{author}{R.~A. {Angryk}},
\newblock \bibinfo{title}{How to train your flare prediction model: Revisiting robust sampling of rare events},
\newblock \bibinfo{journal}{The Astrophysical Journal Supplement Series} \bibinfo{volume}{254} (\bibinfo{year}{2021}) \bibinfo{pages}{23}.
\bibitem[{Chen et~al.(2021)Chen, Kempton, Ahmadzadeh, and Angryk}]{chen2021}
\bibinfo{author}{Y.~Chen}, \bibinfo{author}{D.~J. Kempton}, \bibinfo{author}{A.~Ahmadzadeh}, \bibinfo{author}{R.~A. Angryk},
\newblock \bibinfo{title}{Towards synthetic multivariate time series generation for flare forecasting},
\newblock in: \bibinfo{editor}{L.~Rutkowski}, \bibinfo{editor}{R.~Scherer}, \bibinfo{editor}{M.~Korytkowski}, \bibinfo{editor}{W.~Pedrycz}, \bibinfo{editor}{R.~Tadeusiewicz}, \bibinfo{editor}{J.~M. Zurada} (Eds.), \bibinfo{booktitle}{Artificial Intelligence and Soft Computing}, \bibinfo{publisher}{Springer International Publishing}, \bibinfo{address}{Cham}, \bibinfo{year}{2021}, pp. \bibinfo{pages}{296--307}.
\bibitem[{Galar et~al.(2012)Galar, Fernandez, Barrenechea, Bustince, and Herrera}]{galar2012}
\bibinfo{author}{M.~Galar}, \bibinfo{author}{A.~Fernandez}, \bibinfo{author}{E.~Barrenechea}, \bibinfo{author}{H.~Bustince}, \bibinfo{author}{F.~Herrera},
\newblock \bibinfo{title}{A review on ensembles for the class imbalance problem: Bagging-, boosting-, and hybrid-based approaches},
\newblock \bibinfo{journal}{IEEE Transactions on Systems, Man, and Cybernetics, Part C (Applications and Reviews)} \bibinfo{volume}{42} (\bibinfo{year}{2012}) \bibinfo{pages}{463--484}.
\bibitem[{O'Keefe et~al.(2022)O'Keefe, Sadykov, Kosovichev, Nita, Oria, Sharma, Francis, and Chong}]{okeefe2022}
\bibinfo{author}{P.~M. O'Keefe}, \bibinfo{author}{V.~M. Sadykov}, \bibinfo{author}{A.~G. Kosovichev}, \bibinfo{author}{G.~M. Nita}, \bibinfo{author}{V.~Oria}, \bibinfo{author}{S.~Sharma}, \bibinfo{author}{F.~Francis}, \bibinfo{author}{C.~J. Chong},
\newblock \bibinfo{title}{{Handling Highly Imbalanced Data in Machine Learning Applications}},
\newblock \bibinfo{publisher}{Zenodo}, \bibinfo{year}{2022}. \URLprefix \url{https://doi.org/10.5281/zenodo.6780972}. \DOIprefix\doi{10.5281/zenodo.6780972}.
\bibitem[{{Sadykov} et~al.(2017){Sadykov}, {Kosovichev}, {Oria}, and {Nita}}]{sadykov2017}
\bibinfo{author}{V.~M. {Sadykov}}, \bibinfo{author}{A.~G. {Kosovichev}}, \bibinfo{author}{V.~{Oria}}, \bibinfo{author}{G.~M. {Nita}},
\newblock \bibinfo{title}{{An Interactive Multi-instrument Database of Solar Flares}},
\newblock \bibinfo{journal}{The Astrophysical Journal Supplement Series} \bibinfo{volume}{231} (\bibinfo{year}{2017}) \bibinfo{pages}{6}.
\bibitem[{{Papaioannou} et~al.(2016){Papaioannou}, {Sandberg}, {Anastasiadis}, {Kouloumvakos}, {Georgoulis}, {Tziotziou}, {Tsiropoula}, {Jiggens}, and {Hilgers}}]{Papaioannou2016}
\bibinfo{author}{A.~{Papaioannou}}, \bibinfo{author}{I.~{Sandberg}}, \bibinfo{author}{A.~{Anastasiadis}}, \bibinfo{author}{A.~{Kouloumvakos}}, \bibinfo{author}{M.~K. {Georgoulis}}, \bibinfo{author}{K.~{Tziotziou}}, \bibinfo{author}{G.~{Tsiropoula}}, \bibinfo{author}{P.~{Jiggens}}, \bibinfo{author}{A.~{Hilgers}},
\newblock \bibinfo{title}{{Solar flares, coronal mass ejections and solar energetic particle event characteristics}},
\newblock \bibinfo{journal}{Journal of Space Weather and Space Climate} \bibinfo{volume}{6} (\bibinfo{year}{2016}) \bibinfo{pages}{A42}.
\bibitem[{{Breiman}(2001)}]{Brieman2001RF}
\bibinfo{author}{L.~{Breiman}},
\newblock \bibinfo{title}{{Random Forests.}},
\newblock \bibinfo{journal}{Machine Learning} \bibinfo{volume}{45} (\bibinfo{year}{2001}) \bibinfo{pages}{5--32}.
\bibitem[{Tietz et~al.(2017)Tietz, Fan, Nouri, Bossan, and {skorch Developers}}]{skorch2017}
\bibinfo{author}{M.~Tietz}, \bibinfo{author}{T.~J. Fan}, \bibinfo{author}{D.~Nouri}, \bibinfo{author}{B.~Bossan}, \bibinfo{author}{{skorch Developers}}, \bibinfo{title}{skorch: A scikit-learn compatible neural network library that wraps PyTorch}, \bibinfo{year}{2017}. \URLprefix \url{https://skorch.readthedocs.io/en/stable/}.
\bibitem[{Kingma and Ba(2015)}]{Kingma2015}
\bibinfo{author}{D.~P. Kingma}, \bibinfo{author}{J.~Ba},
\newblock \bibinfo{title}{Adam: A method for stochastic optimization},
\newblock \bibinfo{journal}{CoRR} \bibinfo{volume}{abs/1412.6980} (\bibinfo{year}{2015}).
\bibitem[{{Bobra} and {Couvidat}(2015)}]{bobra2015}
\bibinfo{author}{M.~G. {Bobra}}, \bibinfo{author}{S.~{Couvidat}},
\newblock \bibinfo{title}{{Solar Flare Prediction Using SDO/HMI Vector Magnetic Field Data with a Machine-learning Algorithm}},
\newblock \bibinfo{journal}{The Astrophysical Journal} \bibinfo{volume}{798} (\bibinfo{year}{2015}) \bibinfo{pages}{135}.
\bibitem[{{Sadykov} and {Kosovichev}(2017)}]{sadykov2017b}
\bibinfo{author}{V.~M. {Sadykov}}, \bibinfo{author}{A.~G. {Kosovichev}},
\newblock \bibinfo{title}{{Relationships between Characteristics of the Line-of-sight Magnetic Field and Solar Flare Forecasts}},
\newblock \bibinfo{journal}{The Astrophysical Journal} \bibinfo{volume}{849} (\bibinfo{year}{2017}) \bibinfo{pages}{148}.
\bibitem[{Yeolekar et~al.(2021)Yeolekar, Patel, Talla, Puthucode, Ahmadzadeh, Sadykov, and Angryk}]{Yeolekar2021}
\bibinfo{author}{A.~Yeolekar}, \bibinfo{author}{S.~Patel}, \bibinfo{author}{S.~Talla}, \bibinfo{author}{K.~Puthucode}, \bibinfo{author}{A.~Ahmadzadeh}, \bibinfo{author}{V.~M. Sadykov}, \bibinfo{author}{R.~A. Angryk},
\newblock \bibinfo{title}{Feature selection on a flare forecasting testbed: A comparative study of 24 methods},
\newblock in: \bibinfo{booktitle}{2021 International Conference on Data Mining Workshops (ICDMW)}, \bibinfo{publisher}{IEEE Computer Society}, \bibinfo{address}{Los Alamitos, CA, USA}, \bibinfo{year}{2021}, pp. \bibinfo{pages}{1067--1076}. \URLprefix \url{https://doi.ieeecomputersociety.org/10.1109/ICDMW53433.2021.00138}. \DOIprefix\doi{10.1109/ICDMW53433.2021.00138}.

\end{thebibliography}







\end{document}